\documentclass[12pt]{article}

\usepackage[ascii]{inputenc}
\usepackage{amsmath,amssymb,amsfonts,amsthm}
\usepackage[margin=2.8cm]{geometry}
\usepackage{graphicx}
\usepackage[caption=false]{subfig}
\usepackage[pdftex,bookmarks=false,colorlinks=true,linkcolor=blue,
citecolor=blue,filecolor=black,urlcolor=blue]{hyperref}

\usepackage{lmodern,bm}
\usepackage{latexsym}
\usepackage{longtable}
\usepackage{epsfig}
\usepackage{psfrag}

\begin{document}
\begin{center}
\vspace*{2cm} {\large {\bf The $1\!+\!1$ dimensional Kardar-Parisi-Zhang  equation: more surprises\bigskip\bigskip\\}}
{\large Herbert Spohn}\bigskip\bigskip\\
Physik Department and Zentrum Mathematik,
Technische Universit\"at M\"unchen,
Boltzmannstra{\ss}e 3, 85747 Garching, Germany. 
{\tt spohn@tum.de}
\vspace{5cm}\end{center} 
\textbf{Abstract}.
In its original version the KPZ equation models the dynamics of an 
interface bordering a stable phase against a metastable one. Over 
past years the corresponding two-dimensional field theory has been 
applied to models with different physics. Out of a wide choice, 
the spin-spin time correlations for the Heisenberg chain will be discussed at some length, also
the equilibrium time-correlations of the conserved fields for 1D fluids. 
An interesting recent theoretical advance is the construction of the scale-invariant 
asymptotic theory, the so-called KPZ fixed point. 
\\\\ \\ 
\textit{1.~Introduction}. The ground breaking 1986 contribution of Kardar, Parisi, and Zhang, for short KPZ,  is entitled ``Dynamic scaling of growing interfaces'' \cite{KPZ86}, see also the early reviews \cite{BaSt95,Kr97,Me98}. They study the dynamics of  an interface dividing a stable bulk phase from a metastable one. As such this could be in any dimension. But much of the excitement over the last decade refers to a two-dimensional bulk, hence a one-dimensional interface, see \cite{Ta18} for a recent account. Many interesting advances have also been achieved for  higher dimensions \cite{HaTa15}, but to be concise my contribution will be restricted  to one dimension. The $1\!+\!1$ in the title refers to space-time.

Assuming that the bulk dynamics is not constrained by any conservation laws and relaxes exponentially fast,  KPZ argue the motion of the interface to be governed by the stochastic PDE
\begin{equation}\label{1}
\partial_t h = \tfrac{1}{2}\lambda (\partial_x\,h)^2  + \nu (\partial_x)^2 h +\sqrt{D} \xi\,,
\end{equation}
where  $h(x,t)$ denotes the height for the location of the interface relative to a substrate space $x \in \mathbb{R}$ at  time $t \geq 0$. Possible overhangs are discarded. The nonlinearity arises from the asymmetry between the two phases: at the interface a transition from metastable to stable is fast while the reverse process is strongly suppressed. The Laplacian, $(\partial_x)^2$, models the interfacial tension and the space-time white noise,  $\xi(x,t)$,  quantifies the randomness in transitions from metastable to stable. $\lambda, \nu, D$ are material parameters, following the original KPZ notation. 
Later on it will be convenient to choose space-time units such that $\nu = \tfrac{1}{2}$, $D = 1$.
Eq. \eqref{1} does not satisfy detailed balance. In this sense we study a stochastic system belonging to nonequilibrium statistical mechanics.

To have a concrete physical picture and a better understanding of the approximations underlying 
\eqref{1}, it is illuminating to first consider the two-dimensional ferromagnetic Ising model  with Glauber spin flip dynamics, as one of the most basic model system of statistical mechanics. Its spin configurations are denoted by $\sigma = \{\sigma_{\boldsymbol{j}}, \boldsymbol{j} \in \mathbb{Z}^2\}$ with $\sigma_{\boldsymbol{j}} = \pm 1$.  The Ising energy is 
\begin{equation}\label{2}
H(\sigma) = - \sum_{\boldsymbol{i},\boldsymbol{j} \in \mathbb{Z}^2, |\boldsymbol{i}-\boldsymbol{j}|=1} \sigma_
{\boldsymbol{i}}\sigma_{\boldsymbol{j}} - h \sum_{\boldsymbol{j}\in \mathbb{Z}^2}\sigma_{\boldsymbol{j}}\,,
\end{equation}
where the first sum is over nearest neighbor pairs and the spin coupling is used as energy scale. The flip rate from $\sigma_{\boldsymbol{j}}$ to  $-\sigma_{\boldsymbol{j}}$ is given by 
\begin{equation}\label{3}
c_{\boldsymbol{j}}(\sigma) = \left\{\begin{array}{ll}
1, &\mbox{if $\Delta_{\boldsymbol{j}} H(\sigma) \leq 0$\,,}\smallskip\\
\mathrm{e}^{-\beta \Delta_{\boldsymbol{j}} H (\sigma)},& 
\mbox{if $\Delta_{\boldsymbol{j}} H(\sigma) > 0$\,,}
\end{array}
\right. 
\end{equation}
where $\beta = T^{-1}$ is the inverse temperature and  $\Delta H_{\boldsymbol{j}}$ the energy difference in a spin flip at ${\boldsymbol{j}}$,  more explicitly   $\Delta H_{\boldsymbol{j}}(\sigma) = H(\sigma^{(\boldsymbol{j})}) - H(\sigma)$ where $\sigma^{(\boldsymbol{j})}$
equals $\sigma$ except for $\sigma_{\boldsymbol{j}}$ flipped to $- \sigma_{\boldsymbol{j}}$.
The bulk dynamics has no conservation law and, away from criticality, relaxes exponentially fast. Under Kawasaki dynamics, which conserves magnetization, the interface dynamics to be studied would have very different properties \cite{Sp86}. 

For $h=0$ and $T<T_\mathrm{c}$ the Ising model has two stable phases, the  $+$ and $-$ phase. We start the dynamics with one half-space in the $+$ phase joined to the remaining  half space in the $-$ phase. Then the interface builds up fluctuations, but  stays put on average by symmetry. Observed on a coarse scale the interface is governed by \eqref{1} with nonlinearity 
dropped, i.e. $\lambda = 0$. 
On this basis the interface is predicted to have Gaussian fluctuations with a width increasing as $t^\frac{1}{4}$. Changing the Glauber dynamics to a small $h >0$ makes the $-$ phase metastable and the interface acquires a net motion, which in approximation gives rise to a nonlinearity as in \eqref{1}. By power counting, the nonlinearity dominates and the interface is now predicted to broaden faster, namely as    $t^\frac{1}{3}$. More interestingly, even on large scales the fluctuations turn out to have non-Gaussian 
universal statistics. 

Two simple transformations suggest a very different physical interpretation. \medskip\\
(i) We define the slopes $u(x,t) = \partial_x h(x,t)$. Then  
\begin{equation}\label{4}
\partial_tu + \partial_x\big(-\lambda u^2 - \tfrac{1}{2}\partial_x u -  \xi \big) = 0,
\end{equation}
called stochastic Burgers equation. It is a conservation law with the current being the sum of three physically natural terms, a systematic part nonlinear in the field, a stabilizing
response proportional to the slope, and a noise uncorrelated in space-time. \medskip\\
(ii) The second case employs the Cole-Hopf transformation, 
\begin{equation}\label{5}
Z(x,t) = \mathrm{e}^{\lambda h(x,t)},
\end{equation}
which ``linearizes'' the KPZ equation as
\begin{equation}\label{6}
\partial_t Z(x,t) =  \tfrac{1}{2} (\partial_x)^2Z(x,t)  +\lambda\xi(x,t)Z(x,t)\,.
\end{equation}
$Z(x,t)$ should be viewed as a random partition function. The Laplacian generates a Brownian motion, which we regard as a polymer
in the space-time plane. The polymer is
directed since it cannot bent backwards in time. The directed polymer is subject to an uncorrelated  space-time random potential. The polymer energy is the sum of an elastic energy and a random potential energy.   We have thus transformed to an equilibrium problem with disorder. Note that $ h = \lambda^{-1} 
\log Z$ and the height equals the random free energy of the directed polymer. 
\\\\ 
INTERLUDE: Statistical Mechanics has been very successful in developing tools which have a much wider applicability than originally anticipated.
As to be discussed, the KPZ equation is a good example. Because of the occasion, I allow myself to use as illustration the most fundamental formula 
in our field, namely  the probability distribution 
\begin{equation}\label{7}
Z^{-1} \mathrm{e}^{-\beta H},
\end{equation}
describing the statistical properties of matter in thermal equilibrium. A variant of this formula, and its general significance, was discovered by Boltzmann in 1868, two years after his Ph.D. \cite{B68a}. The formula managed to survive the quantum revolution, it reappears in quantum field theory, in the  theory of dynamical systems, in pattern recognition, and many more areas. Admittedly Boltzmann's papers are not easy to read,  even when mastering the required German language. Fortunately, in the same year he wrote the  ten page article ``L\"{o}sung eines mechanischen Problems'' (Solution of a mechanical problem) \cite{B68b}, which I strongly recommend as a  primary source.
Boltzmann considers a point-particle moving  in the plane subject to the radial potential $V(\boldsymbol{q}) = |\boldsymbol{q}|^{-2} -  |\boldsymbol{q}|^{-1}$, an integrable dynamics. He then adds an elastically reflecting wall located 
along a line not intersecting the origin and argues that,  if the forces are confining, the long-time
statistics of the particle's motion is well modelled by the micro-canonical ensemble on the surface of constant energy, see \cite{G16} for a recent discussion. In the  preceding lengthy article \cite{B68a} Boltzmann considers interacting systems with many degrees of freedom and argues for the micro-canonical ensemble as the proper description of thermal equilibrium.\\\\
\textit{2.~Universality and initial conditions}. Let us solve \eqref{1} with flat initial conditions, $h(x,0) = 0$, and study the height at a given point, say the origin,
as a function of time, i.e. the random function $t \mapsto h(0,t)$. Anticipating height fluctuations of size $t^\frac{1}{3}$ we  expect a systematic drift plus fluctuations,  
\begin{equation}\label{8}
h(0,t) \simeq vt + (\Gamma t)^{\frac{1}{3}}\xi_\mathrm{flat}, 
\end{equation}
to hold for long times. $v$ is the macroscopic growth velocity. The random amplitude, $\xi_\mathrm{flat}$, 
has a universal distribution, while the time scale $\Gamma^{-1}$ collects model specific properties.
For example for the Ising interface discussed above, the distribution of $\xi_\mathrm{flat}$ should not depend on temperature and on additional  next nearest neighbor couplings. 
This universal distribution was first obtained by studying  the polynuclear growth model (PNG) \cite{PS00} and identified as 
\begin{equation}\label{9}
\xi_\mathrm{flat} = \xi_\mathrm{GOE},
\end{equation}
which is the Tracy-Widom distribution of the largest eigenvalue of a $N\times N$ GOE  random matrix in the limit of large $N$ \cite{TW93}. In fact the distribution function of $\xi_\mathrm{GOE}$ is  most easily written in terms of an infinite-dimensional Fredholm determinant \cite{FS07},
\begin{equation}\label{10}
\mathbb{P}(\{\xi_\mathrm{GOE} \leq s\}) =\det( 1- K_s)_{L^2(\mathbb{R}_+)},
\end{equation}
where $K_s$ is the integral operator with kernel $\langle x|K|x'\rangle = \tfrac{1}{2}\mathrm{Ai}(x+x' +s)$, $\mathrm{Ai}$ the standard Airy function. [\,$\mathbb{P}(A)$ is our generic notation for ``probability of the set $A$'']. Later on the same distribution was derived for other growth models \cite{Sa06,WFS17} and eventually also confirmed through the famous Sano-Takeuchi experiment \cite{TaSa12}. So there is little doubt that the asymptotics \eqref{8} together with \eqref{9} holds in great generality.
\begin{figure}[ht!]
\centering
\includegraphics[width=0.7\textwidth]{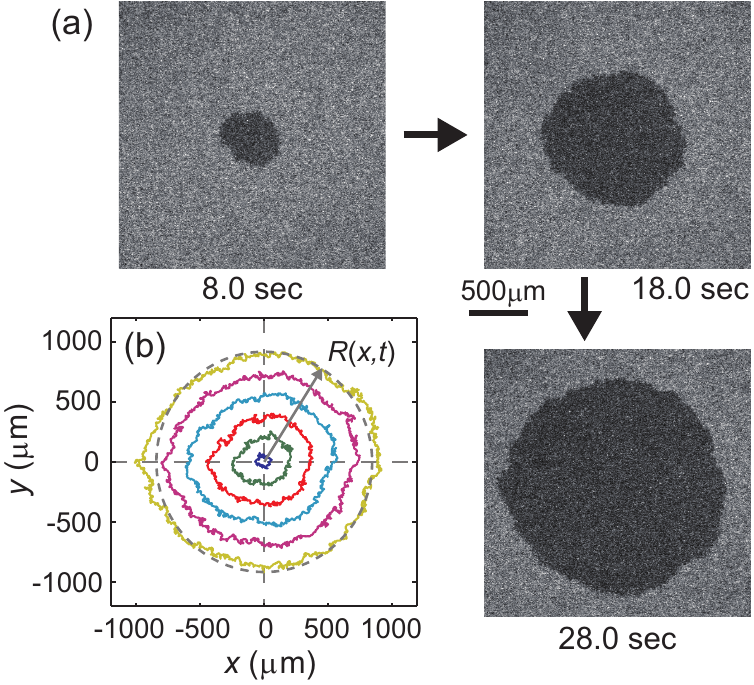}
\caption{Droplet growth in the Takeuchi-Sano turbulent liquid crystal experiment. Shown is light transmission intensity for three time instances. 
Grey is the metastable and dark the stable phase.  $R(x,t)$ is equated with $h(0,t)$ from the theory. }
\label{fig1}
\end{figure}
\begin{figure}[ht!]
\centering
\includegraphics[width=0.85\textwidth]{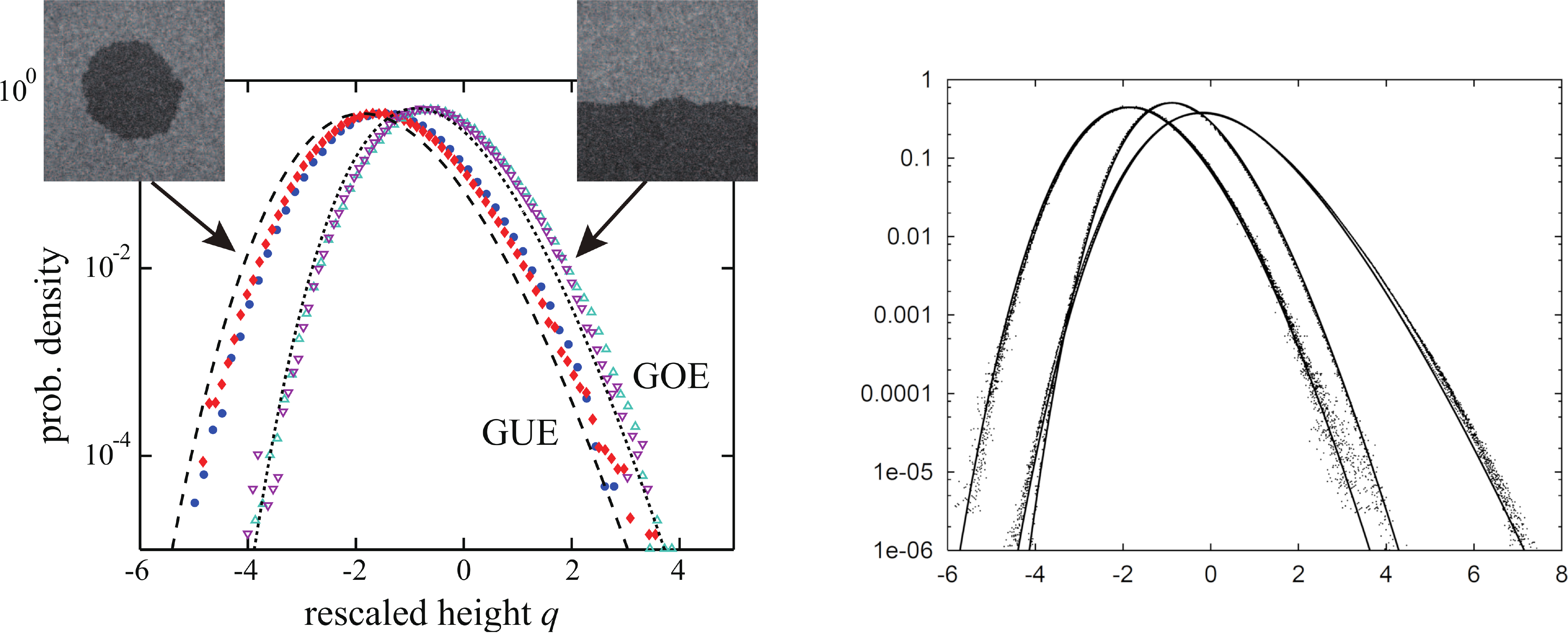}
\caption{Single point height statistics in a logarithmic plot. On the left displayed are data from the Takeuchi-Sano turbulent liquid crystal experiment with curved and flat initial data. 
On the right are shown the exact results from the PNG model. The  right most curve corresponds to Baik-Rains distribution for stationary initial conditions.}
\label{fig2}
\end{figure}

Growing surfaces can be curved, see Fig. 1 for droplet growth. In the stochastic Ising model, one could choose one quadrant in the metastable and the other three quadrants in the stable phase. The corner would then approximately be rounded off into a quarter circle with radius growing proportional to $t$. To model such a set-up with \eqref{1}, instead of flat initial conditions,  we now choose an initial  sharp wedge profile as
$h(x,0) = -(x/a)^2$, $a \ll 1$. Then on average the height function will have a linear shift and a local curvature decreasing as $1/t$, $\langle h(x,t)\rangle \simeq vt - (x^2/a't)$. 
Following the road map \eqref{8}, the universal object are the height fluctuations. They are still of order $t^{\frac{1}{3}}$. However, as a at the time great surprise
\cite{Jo00, PS00}, the distribution
is now modified and given by $\xi_\mathrm{GUE}$,
the Tracy-Widom GUE distribution. For an initially thermally rough interface, i.e. $x \mapsto h(x,0)$ two-sided Brownian motion with
$\langle h(x,0)^2\rangle = \sigma^2 |x|$, one finds yet
another  distribution. In fact it is a one-parameter family of distributions depending on the roughness $\sigma$ of the initial conditions \cite{CFS18}. There is one special value
of $\sigma$ for which such initial conditions are stationary in time. In this case, as discovered Baik and Rains \cite{BR00}, there is an explicit formula for the distribution of the noise amplitude. Thus the emergent picture is richer than anticipated in the early KPZ days. The initial conditions themselves are divided into subclasses, which reflect their behavior 
``at infinity''. For example, if one would locally modify the flat initial conditions, or merely pick subdiffusively growing spatial fluctuations, then the asymptotic behavior is still be given by  \eqref{8}, \eqref{9}. In case of a mixed asymptotic behavior, such as
$h(x,0) = 0$ for $x \leq 0$ and $h(x,0) =  -(x/a)^2$ for $x>0$, the asymptotic fluctuations of $h(0,t)$ are given by  yet another distribution, 
see \cite{C12} for more details.
 
On purpose, we have remained  somewhat vague for which precise model and on which level of rigor an asymptotic limit as \eqref{8} has been established. Under
the header of ``integrable probability'' many results have become available. For this the interested reader is invited to 
consult the cited surveys and original articles. In some cases the scaling limit has been proved  directly for the KPZ equation itself \cite{QS15}. A wide class of results is based on replica solutions to the  KPZ equation \cite{KDP18}. A further powerful approach relies on suitable discretizations of the KPZ equation, some of them being mentioned in Section 5.   
\\\\
\textit{3. Spin correlations for the Heisenberg chain}. If detailed is balance is violated, there is no systematic construction of how to obtain the stationary states. The one-dimensional KPZ equation is an exception. Indeed its stationary measures are of the form
\begin{equation}\label{11}
h_\mathrm{stat}(x) = \rho x +  B(x),
\end{equation}
where $B(x)$ is two-sided standard Brownian motion.  The slope $\rho$ is a free parameter. Since the dynamics is noisy, the system 
attempts to locally reach  stationary. Thus a solution to the KPZ equation will locally look like \eqref{11}. Of course, the more global behavior requires further analysis.
In turn the steady states of the stochastic Burgers equation are $u_\mathrm{stat}(x) = \rho +  \xi_\mathrm{w}(x) $ with $\xi_\mathrm{w}(x)$
normalized white noise in one dimension. One of the available exact results for the KPZ equation \cite{BoCoFeV14}, see also \cite{PrSp04,FS06}, tells us that, setting $\rho = 0$, the stationary two-point function of the Burgers equation is given by 
\begin{equation}\label{12}
\langle u(x,t)u(0,0)\rangle \simeq (\Gamma t)^{-\frac{2}{3}}f_\mathrm{KPZ} \big((\Gamma t)^{-\frac{2}{3}}x\big).
\end{equation}
For the KPZ equation $\Gamma = \sqrt{2}|\lambda|$. But in general it will be some other non-universal coefficient. The scaling function $f_\mathrm{KPZ}$ is tabulated \cite{Prhp},
$f_\mathrm{KPZ}(x) >0$, $f_\mathrm{KPZ}$ is even, and normalized as $\int \mathrm{d}xf_\mathrm{KPZ}(x)=1$. $f_\mathrm{KPZ}$ looks like a Gaussian
but has a somewhat faster fall-off in the tails, more precisely $f_\mathrm{KPZ}(x) \simeq \exp[-0.295 |x|^3]$. 
 
The surprise is that the equilibrium spin-spin time correlation of the spin-$\tfrac{1}{2}$ Heisenberg  chain is well approximated by  $f_\mathrm{KPZ}$
\cite{LZP19}. Currently this is a numerical observation, but  let us explain in more detail.
\begin{figure}[ht!]
\centering
\includegraphics[width=0.9\textwidth]{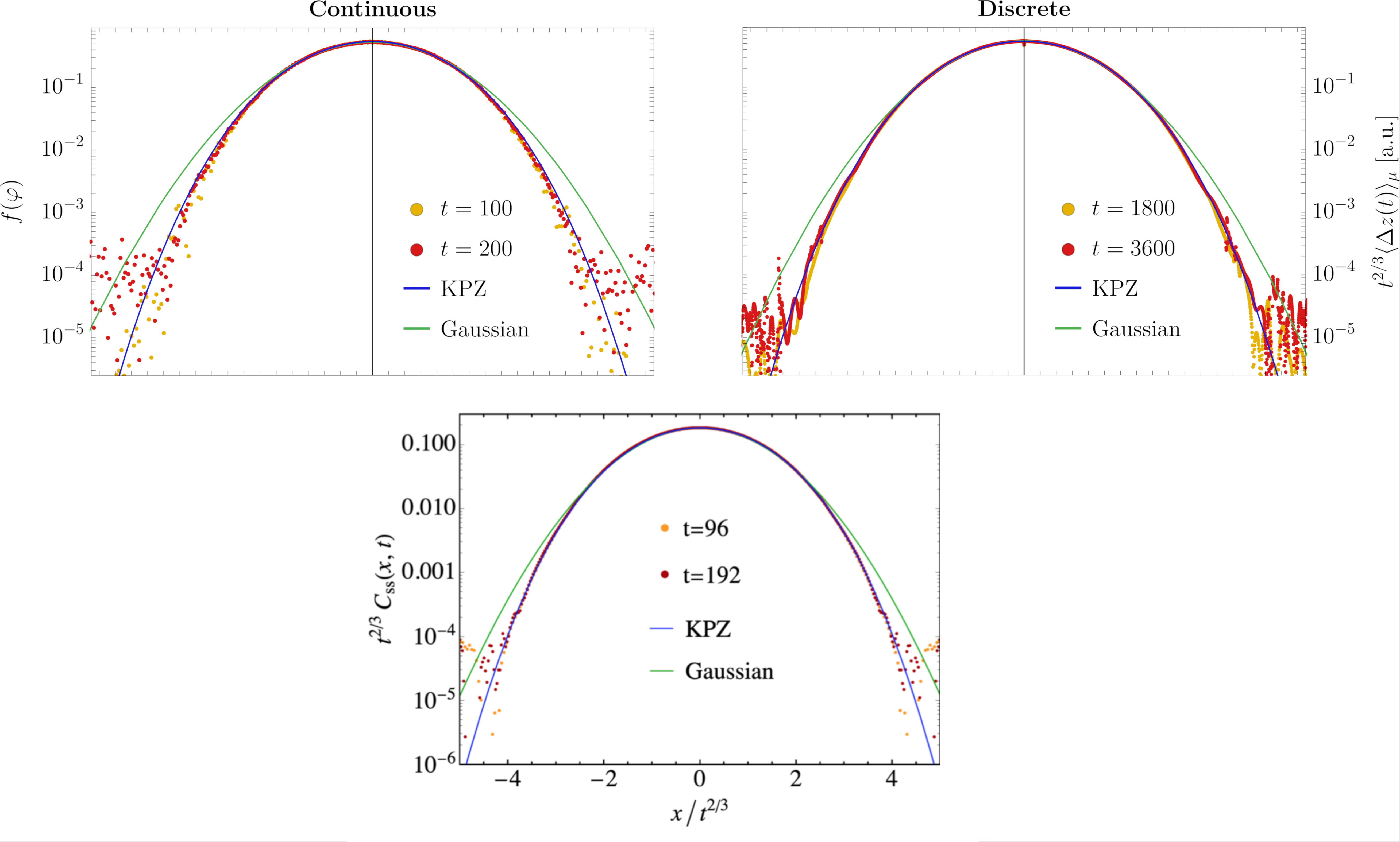}
\caption{Logarithmic plot of the spin-spin time correlation. On the top  displayed are numerical data for the isotropic Heisenberg model at infinite temperature,
left figure continuous time and right figure discrete time, and on the bottom the numerical data for the integrable lattice Landau-Lifshitz spin chain at temperature 1.}
\label{fig3}
\end{figure}
 
The hamiltonian of the Heisenberg spin chain is given by 
 \begin{equation}\label{13}
H = -J\sum_{j\in\mathbb{Z}} \boldsymbol{\sigma}_j\cdot \boldsymbol{\sigma}_{j+1}
\end{equation}
with $ \boldsymbol{\sigma}_j$ the three-vector of Pauli spin matrices at site $j$. The time evolved spin is $ \boldsymbol{\sigma}_j(t) =\mathrm{e}^{-\mathrm{i}Ht }\boldsymbol{\sigma}_j
\mathrm{e}^{\mathrm{i}Ht}$. Using rotational invariance, the quantity of interest is then
\begin{equation}\label{14}
C(j,t) = \langle\sigma^z_j(t)\sigma^z_0(0) \rangle_\beta
\end{equation}
with $ \langle\cdot \rangle_\beta$ denoting thermal average according to the density matrix $Z^{-1} \exp{(-\beta H)}$. Most DMRG simulations are at infinite temperature, $\beta = 0$. But one expects the asymptotics 
not to depend on $\beta$. On the other hand,  the cross-over time does so and at low temperatures it might become impossible to reach the true long time regime.
Numerically apparently more efficient is to adopt a discrete time dynamics, which means
a unitary up-date of pairs of neighboring spins, evenly blocked alternating with pairs oddly blocked. Of course the full SU(3)-invariance and integrability must be preserved \cite{LZP19}. In Fig. 3 top we show numerical results both for continuous and discrete time. In the former case the maximal lattice size is $N = 400$,
time $t = 200$, while in the latter $N= 7200$ and $t = 3600$. As confirmed by Fig. 3  the scaling with $t^{\frac{2}{3}}$ holds with high precision. While indicative this does not yet confirm KPZ behavior. Up to the precision level $10^{-3}$ the fit by a Gaussian seems still to be convincing. But increasing the  level of precision to $10^{-4}$
the fit by $f_\mathrm{KPZ}$ becomes superior. A theoretical analysis is attempted in \cite{DN19}, \cite{B19} and, with a differing perspective, in \cite{M19}.

In the semiclassical limit, the vector of Pauli matrices is replaced by a vector $\boldsymbol{S}_j$ of unit length $|\boldsymbol{S}_j| = 1$ and the 
hamiltonian dynamics is 
governed by
\begin{equation}\label{15}
\frac{d}{dt}\boldsymbol{S}_j = \boldsymbol{S}_j \times \boldsymbol{B}_j, \quad \boldsymbol{B}_j = -\boldsymbol{\nabla}_j H, 
\end{equation} 
known as lattice Landau-Lifshitz spin chain. The interaction is still 
nearest neighbor and isotropic, but has to be chosen such that the dynamics remains integrable. As discovered by Sklyanin
\cite{sklyanin82,sklyanin88}, see also 
Faddeev and Takhtajan \cite{FT07}, 
integrability can be achieved by choosing the interaction 
\begin{equation}\label{16}
h(\boldsymbol{S}_j,\boldsymbol{S}_{j+1}) = -\log (1+ \boldsymbol{S}_j\cdot\boldsymbol{S}_{j+1}). 
\end{equation} 
The corresponding equilibrium state is given by
\begin{equation}\label{17}
\prod_j \big(1 + \boldsymbol{S}_j\cdot\boldsymbol{S}_{j+1}\big)^\beta.
\end{equation} 
For $\beta < 0$ this Boltzmann weight diverges at anti-parallel neighboring spins and can no longer be normalized once $\beta \leq -1$.
Close to that value typically the chain will have long anti-ferromagnetic domains, which slow down the evolution.   A trace of this feature is still present at $\beta = 0$.
Even with a huge number of samples the data are still too noisy to pin down the tail behavior. More stable numerical data are achieved for $\beta = 1$, the   
plots being shown in Fig. 3 bottom \cite{Dhar19}.   $f_\mathrm{KPZ}$ is again confirmed, more or less with the same precision as in the quantum Heisenberg model.
In the numerical simulation of the quantum chain, the employed DMRG evolves directly the reduced density matrix and the plots in Fig. 3 top are based  on averaging over 
a few hundred runs only, which should be compared with $10^6$ samples for the corresponding classical model. 

Clearly, in this context KPZ behavior does not distinguish between classical and quantum. But SU(3) invariance and integrability are crucial. 
Yet preserving integrability, SU(3) can be broken by modifying the $\sigma^z$ coupling, usually denoted by $\Delta$, $\Delta = 1$ being the isotropic case.
Then for $\Delta < 1$ a non-zero Drude weight develops, which vanishes for $\Delta \geq 1$. The spin correlation $C(j,t)$ spreads diffusively
with diverging diffusion constant as $|\Delta - 1| \to 0$ \cite{LZP17}.
\\\\
\textit{4. Equilibrium time correlations for one-dimensional fluids}.
When reflecting on potential physical realizations of the stochastic Burgers equation, the most obvious example are fluids in one dimension, either classical or quantum, where we stick here to the better studied case of classical particles interacting through a  short range potential. 
 In the previous section we insisted on integrability. Now we are so to speak on the opposite side and assume fully chaotic dynamics,
 in the sense that mass, momentum, and energy are the only  locally conserved fields. Obviously the Burgers equation has the short-coming
 of dealing only with a single conservation law.  Well, so why not extend Burgers to several components $u_\alpha(x,t)$, $\alpha = 1,...,n$? Following the KPZ road map one then arrives at 
 \begin{equation}\label{18}
\partial_t u_\alpha + \partial_x \big((A \boldsymbol{u}\,)_\alpha + \boldsymbol{u} \cdot (H^\alpha \boldsymbol{u}) -
(D \partial_x\boldsymbol{u}\,)_\alpha + (B  \boldsymbol{\xi}\,)_\alpha \big)  =0\, ,
\end{equation}
$\alpha = 1,...,n$. $A,H^\alpha,D,B$ are $\boldsymbol{u}$-independent constant $n\times n$ matrices. The fields $\boldsymbol{u}(x,t)$ are fluctuating. In case of a  fluid  we have $n= 3$  and think of them as small deviations from a uniform equilibrium state
at fixed thermodynamic parameters, which are suppressed in our notation. Then the matrix $A$ is obtained from linearizing the
Euler equation. Note that, 
in contrast to a single mode, $A$ cannot be removed by a Galilei transformation. In fact, $A$ determines the crucial peak structure. $H^\alpha$ is a symmetric
matrix obtained from the second order Taylor expansion. $D$ is a phenomenological diffusion matrix satisfying the constraints coming from time-reversal invariance of the fluid. The noise term is Gaussian with mean zero and correlator 
$\langle\xi_\alpha(x,t)\xi_{\alpha'}(x',t')\rangle = \delta_{\alpha,\alpha'} \delta(x-x') \delta(t-t')$, added in the spirit of Landau-Lifshitz fluctuation theory. 
Hence one imposes the fluctuation-dissipation relation $DC +CD^\mathrm{T} = -BB^\mathrm{T}$ with $C$ the static susceptibility matrix, see \cite{Sp14} for more details.
In principle one could retain higher order terms. E.g. $D$ itself might depend on the thermodynamic parameters of the fluid and one may want to 
keep also its  first
 order expansion. By power counting one concludes that the second order Euler term is relevant, while all other higher order terms should not contribute to the long time behavior. 

We now focus on a 1D fluid with the generic hamiltonian
\begin{equation}\label{19}
H_\mathrm{fl} = \sum_j \tfrac{1}{2} p_j^2+ \tfrac{1}{2}\sum_{i\neq j}V(q_i - q_j).
\end{equation} 
$q_j, p_j$ is position and momentum of the $j$-th particle with unit mass and $V$ is a short range, thermodynamically stable potential, $V(x) = V(-x)$.  
The thermal states are labelled by $\beta$, the chemical potential $\mu$, and the mean velocity $\mathsf{u}$. For the background fluid, by Galilei invariance,
we set  $\mathsf{u} = 0$. But the small $\mathsf{u}$ behavior will still be needed to carry out the second order expansion from above. The standard definition for the microscopic fields reads
\begin{eqnarray}\label{20}
&&\mathfrak{n}(x) = \sum_j \delta(q_j - x),\qquad \mathfrak{u}(x) =  \sum_j \delta(q_j - x)p_j,\\\label{21}
&&\mathfrak{e}(x) = \sum_j \delta(q_j - x)\tfrac{1}{2}p_j^2+ \tfrac{1}{2}\sum_{i\neq j}\delta(q_i - x)V(q_i - q_j).
\end{eqnarray} 
To have a more compact notation we set $\vec{\mathfrak{g}} = (\mathfrak{n},\mathfrak{u},\mathfrak{e})$. The comparison with the 3-component Burgers equation is based on the claim that in approximation
\begin{equation}\label{22}
\langle \mathfrak{g}_{\alpha}(x,t) \mathfrak{g}_{\alpha'}(0,0) \rangle_{\beta,\mu}^\mathrm{c} \simeq 
\langle u_{\alpha}(x,t) u_{\alpha'}(0,0) \rangle.
\end{equation} 
Of course the time-dependent microscopic fields, $\mathfrak{n}(x,t),\mathfrak{u}(x,t),\mathfrak{e}(x,t)$, are obtained by evolving $q,p$ according to Newton's equations of motion. On the left the average is with respect to thermal equilibrium at $\beta,\mathsf{u} = 0,\mu$. On the right hand side $\boldsymbol{u}$ is the solution to 
\eqref{18} with random initial conditions corresponding to thermal equilibrium, i.e. $\boldsymbol{u}(x,0)$ is Gaussian, mean zero, and has covariance
$\langle u_{\alpha}(x,0) u_{\alpha'}(x',0) \rangle = C_{\alpha,\alpha'} \delta(x- x')$.
\begin{figure}[ht!]
\centering
\includegraphics[width=0.8\textwidth]{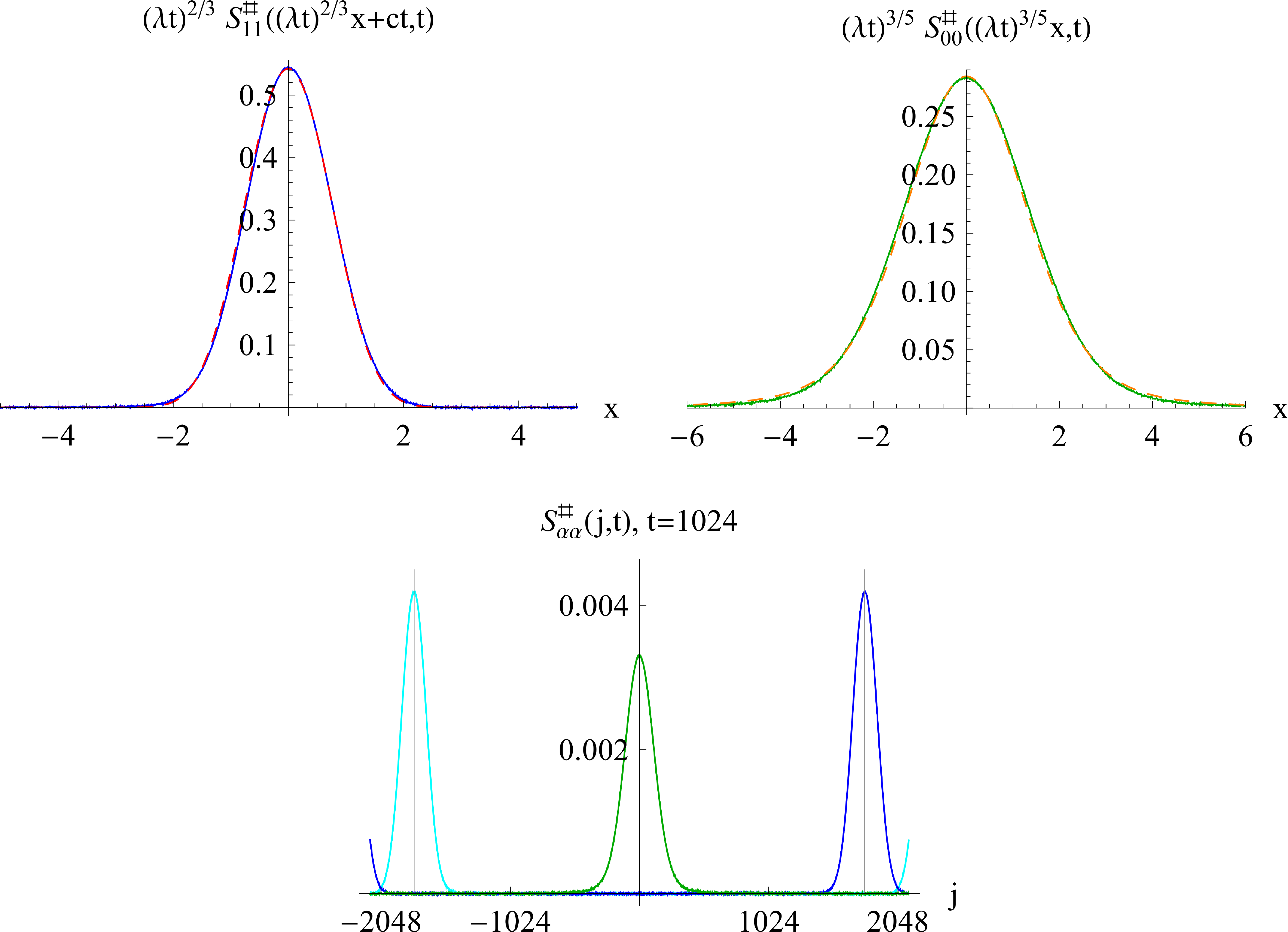}
\caption{Equilibrium time correlations of a fluid with  hard shoulder potential at density $0.8$ and temperature $0.5$. System size is $N = 4096$ and the maximal time $ t= 
1024$. The bottom part shows the standing heat peak and the
linearly in time propagating sound peaks. The sound speed is $c = 2.18$. The comparison with the predictions from nonlinear fluctuating hydrodynamics are displayed on the top, left panel being the sound peak and right panel the heat peak. }
\label{fig4}
\end{figure}

The matrix $A$ has the three eigenvalues $(-c,0,c)$ with $c$ the isentropic speed of sound. Solving \eqref{18} with the said random initial conditions 
and only keeping the $A$-term leads to the three peak structure, $a_-\delta(x+ct) + a_0\delta(t) + a_+ \delta(x - ct)$, where the $a$-coefficients depend on the 
particular correlation considered. Noise and diffusion broaden the peaks to a Gaussian shape with width increasing  as $\sqrt{t}$. As for KPZ,
this is not what is observed numerically. Firstly the peaks broaden superdiffusively and, unexpectedly, the two sound peaks and the central heat peak broaden with distinct exponents
and scaling functions. In approximation the sound peak follows the Burgers equation, i.e. scaling as $ (\Gamma t)^{-\frac{2}{3}}f_\mathrm{KPZ} \big((\Gamma t)^{-\frac{2}{3}}x\big)$, while the central peak is given by the symmetric $\tfrac{5}{3}$ Levy distribution 
\begin{equation}\label{23}
\int_\mathbb{R} \mathrm{d}k \exp\big[- (\Gamma t)|k|^\frac{5}{3} + \mathrm{i}kx\big].
\end{equation} 
As stated, this suggests that all fluids would fall into  a single universality class. But we merely explained the  generic behavior.
Actually, the universality classes are determined by the eigenvalues of $H^\alpha$ as discussed in more detail in \cite{Sp16,PSSS16}.

Numerically, by a Monte Carlo algorithm one samples the thermal initial condition and then evolves according to Newton's equation of motion.
Typically sizes are a few thousand particles and times up to  $t = 4000$ with  density and temperature of order $1$. The average is over $10^7$ initial conditions.  
A much studied model is the FPU chain \cite{DDSMS14}. But solving differential equations is time-consuming and an alternative choice would be a piecewise constant potential \cite{MS14}. Then the collision time vanishes  and in between
two collisions the trajectories are straight lines. In Figs. 4 and 5 we show the results from the simulation of the hard shoulder potential, given by
$V_\mathrm{hs}(x) = \infty$ for  $0\leq |x| \leq \tfrac{1}{2}$, $V_\mathrm{hs}(x) = 1$ for  $\tfrac{1}{2} \leq |x| \leq 1$, and $V_\mathrm{hs}(x) = 0$ for  $1 
\leq |x|$. The shape functions of heat and sound peak agree surprisingly well with the predictions from nonlinear fluctuating hydrodynamics. 
Particularly pronounced are the slow tails of the heat peak, which originate from the cusp in $|k|^\frac{5}{3}$.
This tail is actually cut-off when it reaches the sound peak. Beyond the sound peak the correlations are exponentially small.  In Fig. 5 we display the total current correlation function for momentum and energy. The cross correlations vanish because of distinct symmetry under time-reversal.  The prediction  is $t^{-\frac{2}{3}}$ for both
correlations, but apparently the momentum current correlation has the more rapid convergence. 
\begin{figure}[ht!]
\centering
\includegraphics[width=0.8\textwidth]{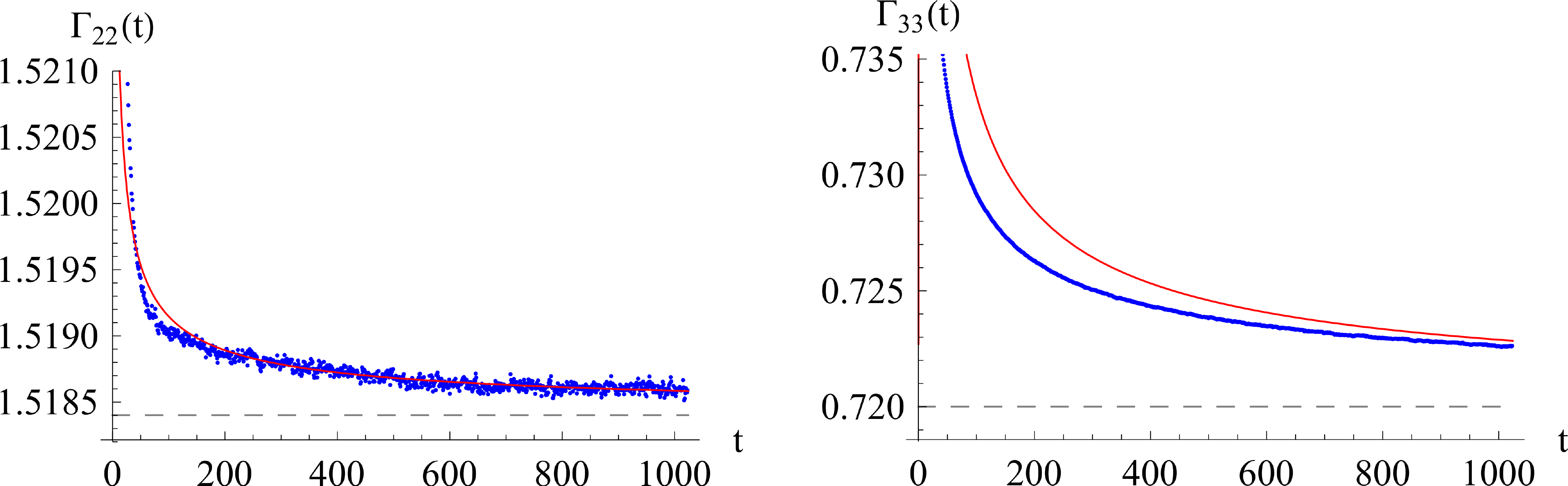}
\caption{Total current-current correlations of a fluid with hard shoulder potential at the same parameters as in Fig. 4. The momentum current correlation  is shown on the left and the
energy current correlation on the right, the fit curve being $t^{-\frac{2}{3}}$ in both cases.}
\label{fig5}
\end{figure}
 
In essence, the multi-component KPZ equation relies on conservation laws.  In particular, it can be used also for stochastic lattice gases \cite{FeSS13,PSS15}.
An interesting case are non-integrable classical spin chains \cite{DDHKMS19}. On face there are only two conserved fields, namely energy and the 
$z$-component of the spin. The corresponding currents vanish in thermal equilibrium, suggesting diffusive spreading, which is  a well confirmed in molecular dynamics simulations \cite{DDHKMS19}. But  for easy plane at low temperatures phase slips are very much suppressed and the field of phase differences becomes essentially conserved.
Now the respective $H^{\alpha}$-matrices  no longer vanish, sound waves persist, and their shape function is reasonably well approximated by $f_\mathrm{KPZ}$.    
 
From this and more simulations there is a lesson to be learned, which is equally important for the one-component KPZ equation. There are considerable cross-over time scales which blur the expected asymptotic behavior. For example the sound peak should be $f_\mathrm{KPZ}$ which  is a symmetric function.
Sound peaks from the simulation have the expected scaling behavior, but one observes a substantial asymmetric distortion which disappears only slowly.  Nonlinear fluctuating hydrodynamics makes also a prediction for the non-universal coefficient $\Gamma$. But the $\Gamma$ observed in Fig. 4 is time-dependent and for the latest time still off by $30\%$. This could also be caused by the employed one-loop computation. One might think that such behavior has its origin in the deterministic time evolution. But the same observations have been reported for one-component fully stochastic systems \cite{SSP16}. 
In the exactly solvable cases, as PNG, single-step, and TASEP, numerically the asymptotic scaling sets in quickly.
But if the  the growth dynamics is modified in a  natural way, as for example the deposition rate depending on the nearby height profile,
then again there is a long time transient regime.\\\\
\textit{5. The KPZ fixed point}.  So far we considered one-point distributions, i.e. the distribution of $h(x,t)$ at a given space-time point.
More generally one might inquire about the scaling behavior of the joint distribution of the height at several points.
One example of interest would be the two-time joint probability $\mathbb{P}(\{h(0,t_1) \leq s_1,h(0,t_2) \leq s_2\})$, see
\cite{ND18,J18}. 
 
Before proceeding, it might be helpful to recall the well-studied example of the 2D ferromagnetic  Ising model at the critical point.
The most basic object is the two-point correlator $\langle \sigma_0 \sigma_{\boldsymbol{j}}\rangle$ and its scaling behavior. Next one studies the 
fully truncated four-point  function  $\langle \sigma_0 \sigma_{\boldsymbol{j} _1}\sigma_{\boldsymbol{j}_2} \sigma_{\boldsymbol{j}_3 }\rangle^\mathrm{c}$,
and higher cumulants. Their scaling behavior can be obtained from a conformal field theory with central charge $c =\tfrac{1}{2}$, which in our present context would be called the (full) Ising fixed point. The advantage is obvious. Universal objects can be computed directly from the fixed point theory without taking recourse to a particular lattice approximation.
 
 We return to KPZ growth and momentarily flat initial conditions $h(x,0) = 0$. As a first step we consider equal time but many spatial points,
 $h(x_1,t),...,h(x_m,t)$ with $x_1 < x_2 <...<x_m$. If the distance $x_{j+1} - x_j$ is very small, then the two heights fluctuate together. On the other side
 if it is too large, the heights fluctuate independently. Nontrivial correlations appear on the scale $t^\frac{2}{3}$, compare with \eqref{12}. In studying the full space-time dependence
 it will be convenient to introduce the dimensionless scale parameter $\epsilon$, $\epsilon \ll 1$, which is chosen such that the spatial points scale as
 $\{\epsilon^{-1}x_j, j = 1,...,m\}$ with $x_j$ fixed and of order 1. Then  time should be scaled as $\epsilon^{-\frac{3}{2}}t$ with $t$ fixed and of order 1.
At such a long time $h(\epsilon^{-1}x,\epsilon^{-\frac{3}{2}}t)$ looks locally like Brownian motion in $x$, hence is of size $\epsilon^{-\frac{1}{2}}$. Thus the universal fixed point should be obtained from
the limit $\epsilon \to 0$ of 
\begin{equation}\label{24}
\epsilon^{\frac{1}{2}} \big(h(\epsilon^{-1}x, \epsilon^{-\frac{3}{2}}t) - \epsilon^{-\frac{3}{2}}vt\big),
\end{equation}
which in the probabilistic literature  is called the  $1:2:3$ scaling, referring to height : space : time. The subtraction corresponds to  a frame co-moving with velocity $v$. 

As known for some
time \cite{BFP07,BFPS07}, the joint distribution of $m$  heights at equal time has a non-degenerate limit,
\begin{equation}\label{25}
\lim_{\epsilon \to 0} \mathbb{P}\big(\{ \epsilon^{\frac{1}{2}} (h(\epsilon^{-1}x_j, \epsilon^{-\frac{3}{2}}t) - \epsilon^{-\frac{3}{2}}vt )\leq s_j, j = 1,...,m\}\big) =  \mathbb{P}\big(\{\mathcal{A}_1(x_j) \leq s_j, j = 1,...,m\}\big).
\end{equation}
On the right hand side appearing are the finite-dimensional distributions of the so-called Airy-1 process. As a stochastic process $\mathcal{A}_1(x)$ is almost surely continuous in $x$, stationary, has $\xi_\mathrm{GOE}$ as one-point distribution, looks locally like a Brownian
motion, and has super-exponentially decaying correlations \cite{BFP08}. The right hand side of expression \eqref{25} can be written in terms of a Fredholm determinant, similarly to \eqref{10}. But
such details can be found in the literature \cite{BFP07}.
A similar formula holds for droplet growth \cite{PrSp03,J05}. Now the additional term $ -\frac{1}{2} x^2/t$ must be added on the right hand side of \eqref{25}
to account for the curvature. As limit one finds the so-called Airy-2 process, which has similar properties as Airy-1, but the power law decay  
$\langle\mathcal{A}_2(x)\mathcal{A}_2(0)\rangle  - \langle\mathcal{A}_2(0)\rangle^2\simeq 1/|x|^2$ for large $|x|$. As before the finite dimensional distributions can be written in terms of a Fredholm determinant.

Such results are based on lattice discretizations of the KPZ equation, the most studied one being the single-step model, equivalently  TASEP.  Now the substrate space is the one-dimensional lattice $\mathbb{Z}$ and the height function $h(j,t)$ is integer-valued, $j \in \mathbb{Z}$, $t \in \mathbb{R}_+$. Single step refers to the constraint $h(j+1,t) - h(j,t) = \pm 1$. Visually it helps to use an interpolating broken line with slope $\pm 1$.
Then flat initial conditions are represented by a zig-zag line, 0 at even and 1 at odd sites. The sharp wedge would be $h(j,0) = |j|$. The growth dynamics has maximal simplicity:  The only allowed transitions are from \scalebox{0.8}{$\diagdown{\hspace{-0.5pt}}\diagup$} to \scalebox{0.8}{$\diagup{\hspace{-0.5pt}}\diagdown$}
and they occur independently with rate 1. More pictorially one considers a large collection of squares of 
side-length $\sqrt{2}$ and rotated by $\pi/4$. One by one squares are randomly deposited
at local minima of the current height profile. For the single-step model the limit \eqref{25} is a proved theorem, but only for flat and sharp wedge initial conditions. For the sharp wedge one exploits a hidden fermionic (= determinantal) structure.
The flat case turned out to be unaccessible for years until Sasamoto \cite{Sa06}
discovered again a determinantal scheme, but on an enlarged space of height functions and giving up positivity of the underlying measure.

The structure of the KPZ fixed point was very recently established by Matetski, Quastel, and Remenik under the $1:2:3$ scaling \eqref{24} using
the single-step model \cite{QM17,MQR17,R19,QR19}. The height evolves according to a Markov process.
Thus one hopes that also the limit $\epsilon \to 0$ remains Markov. Then fixed point means to find out the limiting Markov transition probability on the space of
height functions. The Airy process \eqref{25} is so to speak one family of matrix elements of the transition probability. The initial height is prescribed (either flat or sharp wedge), but the 
height at rescaled time $t$ has a full probability distribution, which is characterized by the right hand side of \eqref{25} for arbitrary $m$ and spatial points $x_j$, $j = 1,...,m$. Missing has been the extension of such a result to ``arbitrary'' initial height profiles. For this one assumes that at time $t=0$ given are a sequence of initial height profiles,
$h_\epsilon(x, 0)$, such that their limit under diffusive scaling holds,
 \begin{equation}\label{26}
 \lim_{\epsilon \to 0} \epsilon^{\frac{1}{2}} h_\epsilon(\epsilon^{-1}x,0)= \mathfrak{h}(x,0).
\end{equation}
Somewhat symbolically the convergence of the transition probabilities means that
\begin{eqnarray}\label{27}
&&\hspace{-20pt}\lim_{\epsilon \to 0}\mathbb{P}\big(\{ \epsilon^{\frac{1}{2}} (h(\epsilon^{-1}x, \epsilon^{-\frac{3}{2}}t) - \epsilon^{-\frac{3}{2}}vt),x \in \mathbb{R}\}\big| \{\epsilon^{\frac{1}{2}} h_\epsilon(\epsilon^{-1}x,0),x \in \mathbb{R}\}     \big) \nonumber\\
&&\hspace{20pt}= \mathbb{P}\big(\{ \mathfrak{h}(x,t),x \in \mathbb{R}\} \big| \{\mathfrak{h}(x,0),x \in \mathbb{R}\} \big).
\end{eqnarray}
While the limit  on the right hand side of \eqref{27} is still a continuous time Markov process, its formal generator is not such a helpful object. Instead, a complete listing of the transition probabilities is provided.

Universal properties of growth processes can be obtained by working directly with the fixed point transition probability. But this is not a simple task at all.
The known cases, as convergence to the Airy processes, can be reproduced, but along very different routes than in the microscopic approach.
Novel cases are still being explored. Just to see, what is involved let us return to the  two-time correlation with flat initial conditions, but now evaluated at the fixed point
\begin{equation}\label{28}
\mathbb{P}(\{\mathfrak{h}(0,t_1) \leq s_1,\mathfrak{h}(0,t_2) \leq s_1\}).
\end{equation}
Just like for a Markov chain with a finite state space, starting from the flat initial condition one has to evolve up to time $t_1$, multiply with the characteristic function of
$\{\mathfrak{h}(0,t_1) \leq s_1\}$, then evolve to time $t_2$ multiply with the characteristic function of
$\{\mathfrak{h}(0,t_2) \leq s_2\}$, and finally average.

The KPZ fixed point structure has little in common with RG as used in critical phenomena of  equilibrium statistical mechanics. Rather  one fully exploits
an initially hidden integrable structure of the single-step stochastic dynamics.\\\\
\textit{6. Even more surprises}.  There are further instances in which the KPZ equation turns out to be central. Merely few cases are listed and
presumably there are more.\\
$-$ Conductance fluctuations in the 2D Anderson model in the localization regime \cite{PSO09, LD16},\\
$-$ Growth of entanglement entropy for the random unitary model \cite{NRVH17},\\
$-$ KPZ scaling for the Kuramoto-Sivashinsky equation \cite{P19}.\\\\
\textbf{Acknowledgements}. I am most grateful to A. Das, C. Mendl, T. Prosen, and K.A. Takeuchi for sharing their numerical plots
and to P.L. Ferrari for a careful reading of the manuscript. In Spring 2019 I was invited as guest professor by the Hausdorff Center for
Mathematics at
Bonn  University, during which time this  contribution was written. I highly appreciate the generous hospitality.

\end{document}